\documentclass[12pt,twocolumn]{aastex6}
\usepackage{amsmath}
\usepackage{amsfonts}
\usepackage{amssymb}
\usepackage{float}
\usepackage{graphicx}
\usepackage{natbib}
\usepackage{rotating}
\begin{document}

\title{Time Domain Filtering of Resolved Images of Sgr A*}

\author{Hotaka Shiokawa\altaffilmark{1}, Charles F. Gammie\altaffilmark{2}, and Sheperd S. Doeleman\altaffilmark{1}}
%\author{Hotaka Shiokawa\footnotemark{1}, Charles F. Gammie\footnotemark{2}, and Sheperd S. Doeleman\footnotemark{1}}
%\author[1]{Hotaka Shiokawa}
%\author[2]{Charles F. Gammie}
%\author[3]{Sheperd S. Doeleman}

%\footnotemark[1]{Harvard-Smithsonian Center for Astrophysics, 60 Garden Street, Cambridge, MA 02138}
%\foornotemark[2]{Department of Physics, University of Illinois, 1110 West Green Street, Urbana, IL, 61801}
\affil{\altaffilmark{1}Harvard-Smithsonian Center for Astrophysics, 60 Garden Street, Cambridge, MA 02138,\\
\altaffilmark{2}Department of Physics, University of Illinois, 1110 West Green Street, Urbana, IL, 61801}
%\altaffiltext{1}{Harvard-Smithsonian Center for Astrophysics, 60 Garden Street, Cambridge, MA 02138}
%\altaffiltext{2}{Department of Physics, University of Illinois, 1110 West Green Street, Urbana, IL, 61801}

\begin{abstract}

The goal of the Event Horizon Telescope (EHT) is to provide spatially
resolved images of Sgr A*, the source associated with the Galactic Center
black hole.  Because Sgr A* varies on timescales short compared to an
EHT observing campaign, it is interesting to ask whether variability
contains information about the structure and dynamics of the accretion
flow. In this paper, we introduce ``time-domain filtering", a technique to filter time fluctuating images with specific
temporal frequency ranges, and demonstrate the power and usage of the
technique by applying it to mock millimeter wavelength images of Sgr A*.
%Here we explore the effects of filtering mock
%observational image data by its power in time variability for specific
%temporal frequency ranges (time-domain filtering).
The mock image data is generated from General Relativistic
Magnetohydrodynamic (GRMHD) simulation and general relativistic
ray-tracing method. We show that the variability on each line of sight is
tightly correlated  with a typical radius of emission. This is because
disk emissivity fluctuates on a timescale of order the local orbital
period. Time-domain filtered images therefore reflect the model dependent
 emission radius distribution, which is not accessible in time-averaged
images.  We show that, in principle, filtered data have the power to
distinguish between models with different black hole spins, different
disk viewing angles, and different disk orientations in the sky.

\end{abstract}

\section{Introduction}\label{sec_introduction}

The Event Horizon Telescope (EHT) is a global very long baseline
interferometry (VLBI) network that will soon produce time- and
space- resolved images of the hot plasma close to the event horizon of
the supermassive black holes at the center of M87 and the Milky Way
\citep{Doeleman.et.al.2009b}.  Already it is clear that both sources
have structure on event horizon scales \citep{Doeleman.et.al.2008,
Doeleman.et.al.2012}, and that in Sgr A* the source varies
on short timescales in both total and polarized intensities
\citep{Fish.et.al.2011, Johnson.et.al.2015b}.  When fully deployed,
the EHT promises to offer higher sensitivity, greater dynamic range, and
greater angular resolution.  This may enable EHT to image the ``photon
ring'', a surface brightness feature that appears on lines of sight
passing close to the photon orbit;
given the accurately measured ratio of black hole mass to distance
\citep[see][]{Psaltis.et.al.2015,Johannsen.et.al.2016}, the angular
radius of the photon ring is uniquely predicted by general relativity
\citep{Bardeen.1973,Luminet.1979}.

At the same time, theoretical developments in modeling black hole
accretion and interpreting EHT data are progressing at a very rapid
pace.  In addition to stationary phenomenological models of black hole
accretion flows \cite[e.g.][]{Narayan.Yi.1994, Broderick.et.al.2009,
Broderick.et.al.2016}, General Relativistic, ideal Magnetohydrodynamics
(GRMHD) codes are now widely available to model time-dependent accretion
flows and jets \citep[e.g.][]{Gammie.et.al.2003, DeVilliers.et.al.2003,
Noble.et.al.2009,  Moscibrodzka.et.al.2014, White.et.al.2016}.
Some models \citep[e.g.][]{Chandra.et.al.2015, Ressler.et.al.2015,
Foucart.et.al.2016} go beyond the ideal fluid approximation and
account for the collisionless nature of the accreting plasma, including
variations in the ratio of ion to electron temperature.  Widely available
relativistic radiative transfer schemes \citep[e.g.][]{Noble.et.al.2007,
Dolence.et.al.2009, Shcherbakov.McKinney.2013, Schnittman.et.al.2013,
Chan.et.al.2015b, Dexter.2016, Narayan.et.al.2016, Gold.et.al.2017}
can, given accurate information about the electron distribution function,
now produce mock images, polarization maps, and spectra of model accretion
flows.

Still, there are obstacles to realizing the full potential of EHT.
Three of the most interesting challenges for Sgr A* are (1) distortion
of the images by electron scattering in the interstellar medium;
(2) theoretical uncertainties related to the structure and dynamics
of the accretion flow; (3) interpreting the time-dependence of the
source, since the intrinsic variability timescales are comparable to
or smaller than the on-source integration time \citep{Lu.et.al.2016,
Medeiros.et.al.2016.arXiv, Medeiros.et.al.2017}.

A large number of pan-chromatic campaigns including radio,
Near-Infrared (NIR), mm/sub-mm, and X-ray have analyzed
the Sgr A* light curves in the context of characteristic timescale
\citep[e.g.][]{Do.et.al.2009,Meyer.et.al.2009,Dexter.et.al.2014} and
flaring emission \citep[e.g.][]{Baganoff.et.al.2001, Genzel.et.al.2003,
Hornstein.et.al.2007, Eckart.et.al.2008, Marrone.et.al.2008,
Porquet.et.al.2008, DoddsEden.et.al.2009, YusefZadeh.et.al.2009,
Trap.et.al.2011, Haubois.et.al.2012, Neilsen.et.al.2013}. While
those studies have revealed the existence of a complex time-variable
emission mechanism likely caused by a magnetized turbulent accretion
flow \citep[e.g.][]{Bower.et.al.2005}, a plausible interpretation
can be obtained only through the time- and space- resolved EHT
observations. Previous VLBI observations have shown evidence of
structural \citep{Fish.et.al.2016} and magnetic field configuration
\citep{Johnson.et.al.2015b} variabilities with a limited number
of baselines. Forthcoming EHT will provide far more information
on the spatial structure of Sgr A*.

It is important to prepare a set of tools to extract information of
 the underlying dynamics from the time variability since intrinsic
variability in EHT observations of Sgr A* is likely. Here, we introduce
 a ``time-domain filtering'' technique that produces images
 of the temporal power in an arbitrary frequency range. The technique
 is in principle applicable to any kind of time fluctuating images,
 and this paper is aimed at demonstration of the usage and power of
the technique by applying it to mock EHT observations.
%Since intrinsic variability in EHT observations of Sgr A* is likely,
%this paper explores what information might in principle be extracted
%from that variability.
We explore what information might be extracted from that variability
using our technique.  Does variability convey information about
turbulence in the accretion flow?  Can this information be extracted
from even ideal mock data?
%Here we introduce a ``time-domain filtering''
%technique that allows us to create images of the temporal power in
%an arbitrary frequency range.
We do not attempt to generate realistic simulated data that includes the
effects of instrumental and atmospheric noise, weather, Earth rotation,
 and interstellar scattering.  Instead we consider time-dependent,
``bare'' mock images of the simulations. We ask (1) how is the frequency
of observational image fluctuations linked to conditions in the emission
region in the disk? (2) does time-domain filtering of the images permit
 one to distinguish between different models?

The plan of the paper is as follows.  \S 2 describes time-dependent
dynamical and radiative models of the accretion flow.  \S 3 presents
method and results of the image filtering.  \S 4 discusses the link
between the image variability and the emission region in the disk,
and \S 5 summarizes the results.

\section{Simulation}

To investigate the effects of time-domain filtering, we run an ideal
 GRMHD simulation of a black hole accretion flow with a small mean
magnetic field.  The flow is assumed radiatively inefficient, so cooling
is negligible.  This is well justified for Sgr A* \cite{Dibi.et.al.2012}.
We then estimate the emergent radiation using a relativistic radiative
transport code.

\subsection{Accretion Disk Simulation}

We evolve the accretion flow using the conservative GRMHD code {\tt
harm3d} \citep{Noble.et.al.2006, Noble.et.al.2009}.  The simulation
starts from a constant angular momentum equilibrium torus
\citet{Fishbone.Moncrief.1976} that is perturbed by a weak magnetic
 field to seed the growth of the Magnetorotational Instability
(MRI). The disk is integrated in Kerr-Schild spacetime in modified
spherical-polar coordinates, with a logarithmically scaled radial grid.
 The computational domain runs from within the event
 horizon to 240$GM_{\rm{BH}}/c^2$ in radius and extends over a full
$2\pi$ radians in azimuth.  The radial and poloidal boundary condition
are set to outflow and reflective boundary, respectively. The simulation
is terminated after 14000$GM_{\rm{BH}}/c^3$, which is long compared to
the timescale for the MRI to reach saturation but short compared to the
 accretion timescale for the torus.  Our initial conditions have only
a small net vertical magnetic flux through the disk, and so produce a
 Standard and Normal Evolution (SANE, \citealt{Narayan.et.al.2012b})
 disk model with short timescale variability that is qualitatively
consistent with observational results \citep{Chan.et.al.2015a}.

We run 2 models with different black hole spin: a rapidly spinning
black hole with $a=0.9375$ (hereafter $0.94$) with resolution
$260{\times}192{\times}128$ (in radius, colatitude, and longitude
respectively), and a nonrotating black hole ($a = 0$) with resolution
$144{\times}144{\times}144$. The difference in resolution arises because
of computational limitations.  One might worry that the difference would
produce differences in time variability properties, but this appears
not to be the case; as one indication, the azimuthal correlation length
of the emissivity, which is tightly correlated to the observational
variability as will be discussed in \S\ref{sec_turbulence}, differs by
only 10\% on average at $r<12GM_{\rm{BH}}/c^2$ (azimuthal correlation length of fluid variables
is discussed in \citealt{Shiokawa.et.al.2012}), while the Innermost Stable Circular Orbit
(ISCO) orbital frequency differs by a factor of $\sim 4$; 34 minutes
and 9 minutes for $a=$0 ($R_{\rm{ISCO}}=6GM_{\rm{BH}}/c^2$) and 0.94 ($R_{ISCO}\sim
2.04GM_{\rm{BH}}/c^2$), respectively. This suggests that the artificial error introduced
by the resolution difference has only a minor effect on the observational
variability in our radiative models.

Hereafter, we describe both the length unit $GM_{\rm{BH}}/c^2\sim
6.64{\times}10^{11}$ cm and time unit $GM_{\rm{BH}}/c^3\sim 22$ seconds
as $M$ by setting $GM_{\rm{BH}} = c = 1$ unless otherwise indicated.

\subsection{Radiative Model}

The next step is to construct time-variable images of the model.
The total duration of our model is $2500M$ ($\sim 15$ hrs for Sgr A*)
taken from the last part of our accretion disk simulation.  We use the
ray-tracing method of \citet{Noble.et.al.2007,Noble.et.al.2009} that
integrates the transfer equation along null geodesics, incorporating
synchrotron emission and absorption.  This leads to a grid of
intensities (pixels) on a ``camera'' at Earth's distance from Sgr A*.
We produce images at $\lambda = 1.3$ mm, where EHT will observe.

In modeling Sgr A* it is common to use a ``fast light'' approximation,
in which the emergent radiation is calculated on individual time slices
and changes in the fluid on the light crossing time are ignored.  But
because we are interested in intensity fluctuations on timescales
comparable to the light crossing time, and plasma close to the event
horizon is moving at relativistic speed, it is necessary to to use a
full ``slow light'' treatment in which the geodesics are integrated
through an evolving flow.  To do this, we update the background snapshot
of the disk every 0.5$M \sim 11$ seconds, as measured by a distant
observer. A photon is advanced by $dx^i$ along its geodesics at each time
step $dx^0=dt=0.5M$ following the relation
\begin{equation}
\frac{dx^i}{dt} = \frac{dx^i}{d\lambda} \bigg/ \frac{dt}{d\lambda}=\frac{k^i}{k^0}.
\end{equation}
Here $\lambda$ is the affine parameter (not the wavelength; the difference
should be clear from the context) and $k^{\mu}$ is the wave four-vector.

%[HS: there is a new paper, Boehle et al. 2016, with slightly improved
%values: 4 x 10^6 and 7.9 kpc.  I don't think we should change the
%calculation, but it would be good to cite this as "see also Boehle et al..."]
The model parameters are the mass of the black hole, the distance
to Sgr A*, the ratio of proton to electron temperature $R \equiv
 T_{\rm{p}}/T_{\rm{e}}$, the accretion rate $\dot{M}$, and the
 inclination angle.  The mass and distance to Sgr A* are set
to $4.5{\times}10^6M_{\odot}$ and $8.4$ kpc, respectively
\citep{Ghez.et.al.2008} (see also \citet{Boehle.et.al.2016} for
the recent estimate).  There are multiple combinations of $R$ and
 $\dot{M}$ that are broadly consistent with the data \citep[see,
 e.g.][for parameter-fitting exercises]{Moscibrodzka.et.al.2009,
 Broderick.et.al.2011, Moscibrodzka.et.al.2014}.  Here we simply fix
$R = 3$ and adjust $\dot{M}$ so that the time-averaged $1.3$mm flux
matches the observed 3.6 Jy \citep{Bower.et.al.2015}.  Finally, for the
 inclination we explore 2 viewing angles: a face-on ($i=2^{\circ}$) and
 edge-on ($i=90^{\circ}$) view of the disk. Since the main purpose of
this paper is to introduce the time-domain filtering technique, we
only examine 4 models, i.e. edge-on and face-on views for
spin=0 and 0.94, to convey a rough sense of how results depend on 
model parameters.

\begin{figure*}[ht!]
%\begin{center}
%\plotone[width=160mm]{flux_snap.pdf}
\plotone{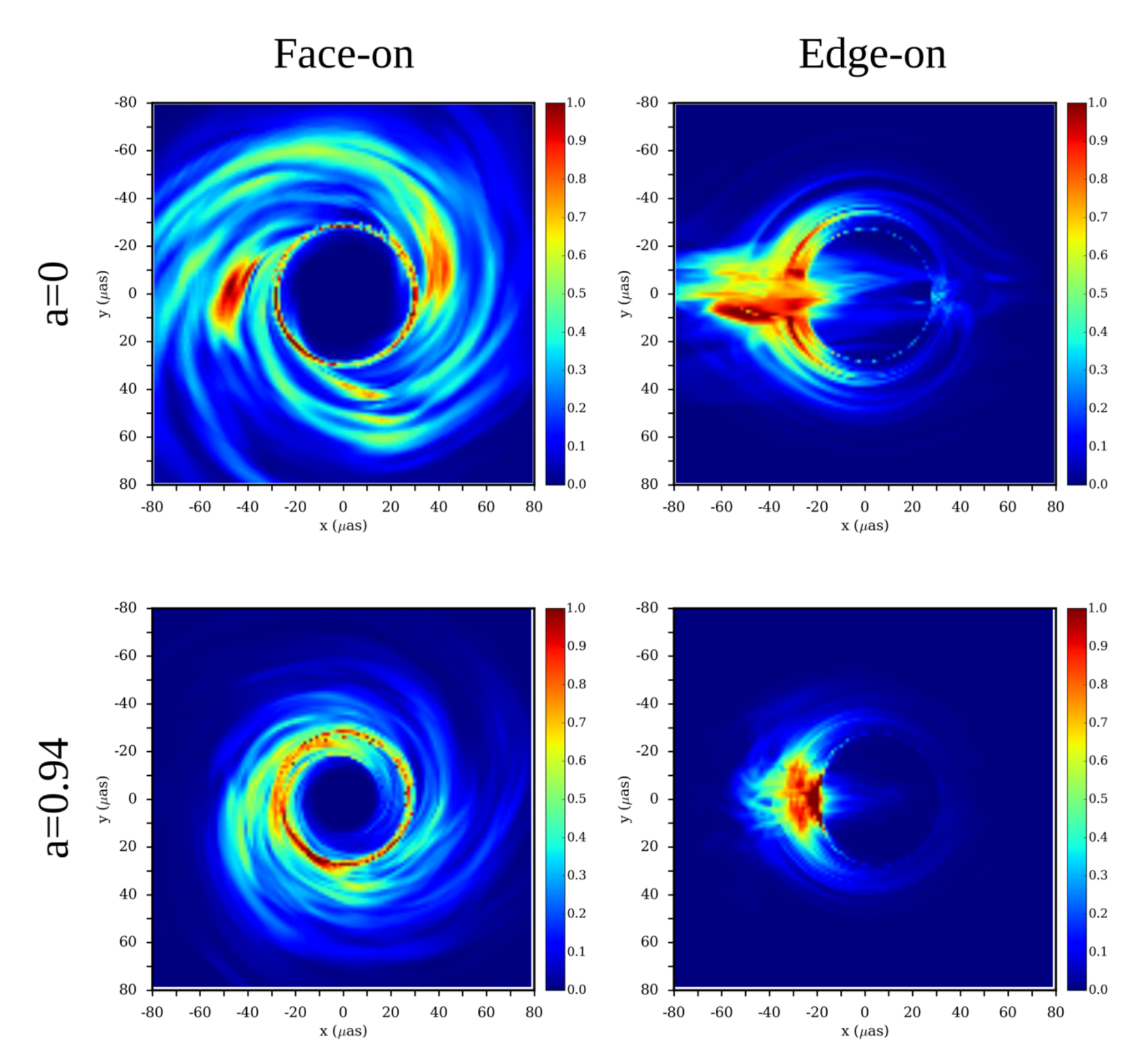}
%\end{center}
\caption[]{Snapshot of ray-traced images at 1.3 mm for the $a=0$ (upper
row) and $a=0.94$ (lower row) simulations. The left and right columns
are the face-on and edge-on view images, respectively.  The resolution
of the images is 128x128. The color scale is linear and normalized.}
\label{flux}
\end{figure*}

Figure \ref{flux} shows ray-traced images of our disk models ($a=0$ and
$0.94$) for face-on and edge-on views of one time slice.  The field of
view (FOV) and image resolution are $30{\times}30M=160{\times}160\,\mu
as$ and $128{\times}128$, respectively, which gives a pixel size of
$1.25{\times}1.25\,\mu as$. In the edge-on image, the left side of the
images is brighter due to Doppler beaming as the plasma moves toward the
camera on the approaching side of the disk. The innermost of the bright
arcs in the edge-on images and the bright rings of the apparent radius
$\sim25-30 \,\mu as$ in the face-on images are the so-called ``photon
ring". They are emission from slightly outside the photon orbit in
the equatorial plane, where $R_{\rm{ph}}=3M (1.43 M)$ for $a=0 (0.94)$
respectively. The apparent size of the photon orbit is increased
by gravitational lensing \citep{Bardeen.1973}.
%In the face-on images,
%the photon ring of radius $\sim25-30 \,\mu as$ is from light rays whose
%geodesics pass through the equatorial plane of the disk multiple times.

\subsection{Emission Radius}\label{sec_emission_radius}

%[HS: we should either use s or lambda for the affine parameter, but not both]
It is convenient to define a characteristic location of emission along
each ray, since our hypothesis is that variability along a ray is
tightly correlated with its point of emission.  Suppose $\lambda$ is the
affine parameter along a geodesic. Then
\begin{equation}
\bar{\lambda} \equiv \frac{\int \lambda \left( \frac{\mathrm{d}I}{\mathrm{d}\lambda} \right)^2 \, \mathrm{d}\lambda}{\int \left( \frac{\mathrm{d}I}{\mathrm{d}\lambda} \right)^2 \, \mathrm{d}\lambda}
\end{equation}
where $I(\lambda)$ is specific intensity. We define the point at $\bar{\lambda}$
along the ray as ``emission point", $\vec{x}(\bar{\lambda})=\vec{x}_{emiss}$;
$r(\bar{\lambda}) \equiv r_{emiss}$ is the characteristic ``emission radius'' and $\theta(\bar{\lambda}) \equiv \theta_{emiss}$
is the characteristic emission latitude.  The bright regions in our model images
have $\theta_{emiss}\sim \pi/2$, i.e. the disk's equatorial plane.

\begin{figure*}[ht!]
%\begin{center}
%\plotone[width=160mm]{emiss_r.pdf}
\plotone{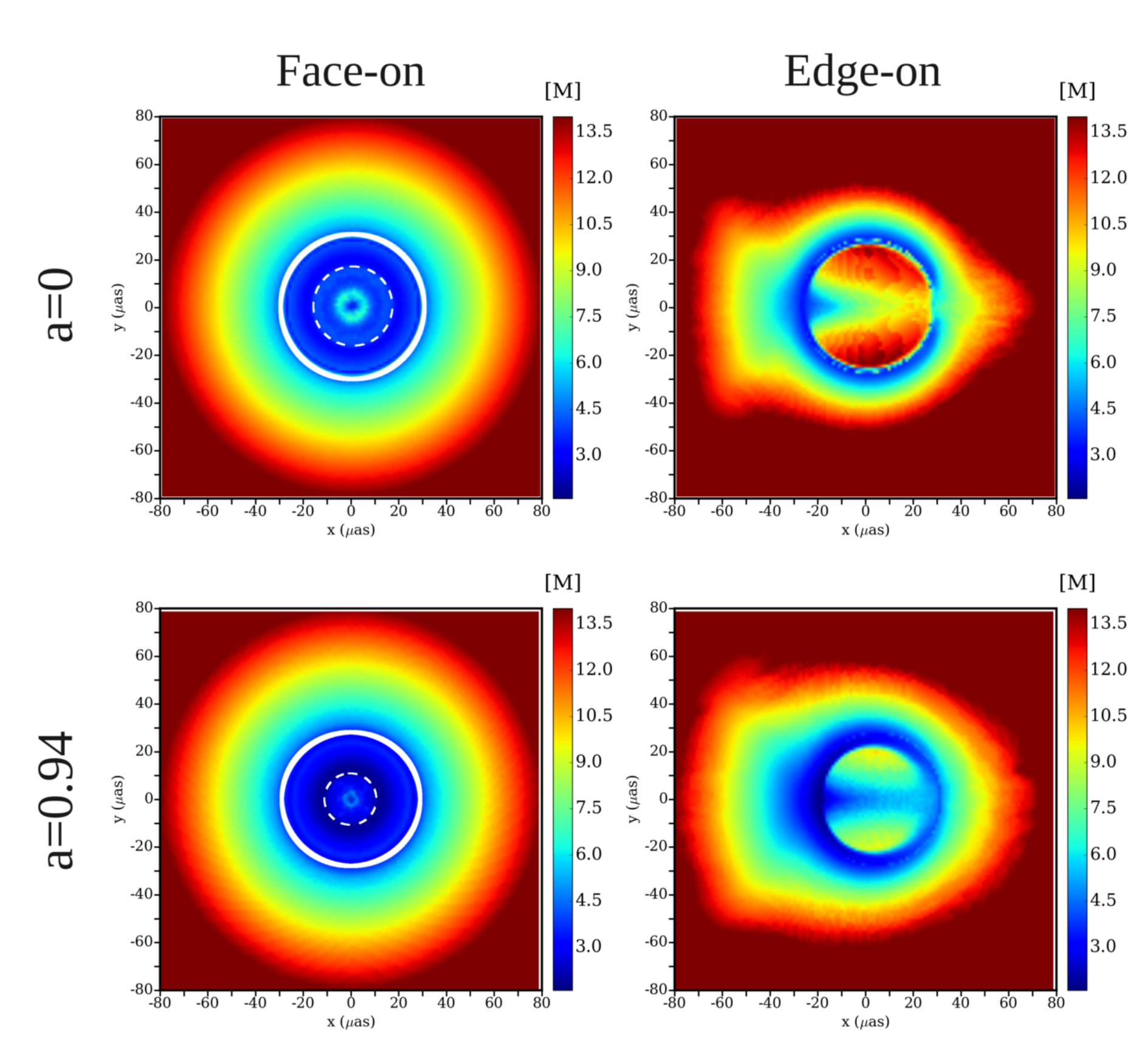}
%\end{center}
\caption[]{Emission radius, $r_{emiss}$, in $M$ for the $a=0$ (upper
row) and $a=0.94$ (lower row) simulations. The left and right columns
are the face-on and edge-on view images, respectively. For the face-on
images, the region within the dotted white ring is for a low emission from
an extended jet region and the $r_{emiss}$ value is not meaningful there.
The solid white ring indicates location of the photon ring where the use of our $r_{emiss}$
definition is inadequate (and therefore hidden by the line) for the face-on images.}
\label{emiss_r}
\end{figure*}

Figure \ref{emiss_r} shows $r_{emiss}$ for the face-on
and edge-on views of both spins.  For the edge-on view, the inner edge
of the photon ring has the smallest $r_{emiss}$ (the blue
rings in Figure \ref{emiss_r}). Most of the emission of the rings is from
the side and behind the black hole, while the region inside the rings
is dominated by emission from gas in front of the black hole; these
geodesics extend into the horizon rather than going around the hole.
Although the shadow angular radius is insensitive to spin, the range of
$r_{emiss}$ that light up the photon ring is strongly dependent on spin:
the minimum $r_{emiss}$ in the equatorial plane for the $a=0$ model is
$\sim 4.5M$ and becomes $\sim 8M$ as the image position moves $20\,\mu
as$ outward, while the same measurement gives $2M$ and $6M$ for the
$a=0.94$ model.  This introduces a difference in image variability for
the two spin models, as discussed in the next subsection.

For the face-on view, emission radius monotonically increases outward
from the center of the images except the region inside the white
dotted ring. There, a very low emission from extended regions outside
the disk (jet) dominates.  There is a noticeable difference between the different
spin models inside the photon rings, indicated by the solid white rings,
where $r_{emiss}$ is smaller by $\sim 1.5-2M$ for the $a=0.94$ model
for a given radius from the image center. The use of our $r_{emiss}$
definition in the photon ring is inadequate for the face-on view since
light rays penetrate the disk multiple times at different radii as
they go around the black hole.

In practice the total intensity in a single pixel is built up from an
extended region around its emission point.  If this emission region
is small, the time variability of the intensity reflects dynamics
at the emission point well.  If the emission region is large, the
variability is built up from variations at multiple locations and it is
difficult to extract information from the light curve.  Nevertheless,
a comparison of Figure \ref{emiss_r} with time-domain filtered images
that are presented in \S\ref{sec_filtering} (Figures \ref{psp_map_edge}
and \ref{psp_map_face}) shows a strong correlation between the emission
radius and time variability in our model images.
%%%%%%%%%%%%%%%%%%%%%%%%%%%%%%%%%%%%%%%%%%%%%%%%%%%%%%%%%%%%%%%%%%%%%%%%%%%%

\section{Time Variability}

Is there any feature of the observed emission that constrains the spin
 of the black hole?  Although the brightest region in the image is
more extended for the $a = 0$ models in Figure \ref{flux}, image size
depends on other model parameters (especially $R$, which can be adjusted
until the image size matches the observed value) and does not in itself
constrain the spin. The size of the photon ring is very weakly dependent
on spin (spin simply moves the centroid of emission at first order in
$a$) and thus a constraint on spin will be difficult to extract from
photon ring imaging alone.

Is it possible to use time-domain information to constrain the spin?
A naive argument suggests that it might be: in our SANE models,
 the observational variability originates from the orbital motion
of turbulent structure in the disk as is discussed below. The
variability timescale is thus tied to the orbital timescale, which in
turn depends on $r_{emiss}$. For example, emission from the brightest
regions in the edge-on views in Figure \ref{flux} are generated at
$\sim R_{\rm{ph}}<R<2R_{\rm{ISCO}}$, where the corresponding orbital
periods for $a=0$ and $a=0.94$ are 6-20 minutes and 12-96 minutes,
respectively. We explore below if this relationship between emission
radius and orbital period may be detectable in EHT data sets.

\subsection{PSD of Total Flux Variability}\label{sec_totalflux}

We begin by examing the power spectrum density (PSD) of the total
(source integrated) mm flux. The PSD is constructed as follows.  For a
light curve $L_n$ where $0\leq n \leq N-1$ is the index of the measured
point in time and $N$ is the total number of points, its discrete
Fourier transform is
\begin{equation}\label{Fourier}
\tilde{L_s} = \sum_{n=0}^{N-1} w_n L_n e^{2\pi isn/N}
\end{equation}
where $-N/2 \leq s \leq N/2$ is the index for the temporal frequency
and $w_n$ is a Hamming window used to reduce the edge artifacts.
Then the PSD is defined as
\begin{equation}\label{psd_unnorm}
P_s \equiv \tilde{L_s} \tilde{L_s^{*}} / W
\end{equation}
where $W=N \sum_{n=0}^{N-1} w_n^2$ \citep{Press.et.al.2002}.
%We also
%define the normalized PSD
%\begin{equation}\label{psd_norm}
%\bar{P}_s \equiv \tilde{L_s} \tilde{L_s^{*}} / W\langle L^2 \rangle
%\end{equation}
%for later use.
We divide each light curve into 4 segments and average
the resultant PSDs where $N\sim 1200$ per segment in our operation. The
duration of each segment is $\sim 3.5$ hours (time resolution of our
data set is $0.5M\sim 11$ seconds) which is close to the integration
time of realistic EHT observation. Averaging over segments also improves
the signal-to-noise ratio.

\begin{figure}[ht!]
%\begin{center}
\hspace*{-7mm}\includegraphics[width=100mm]{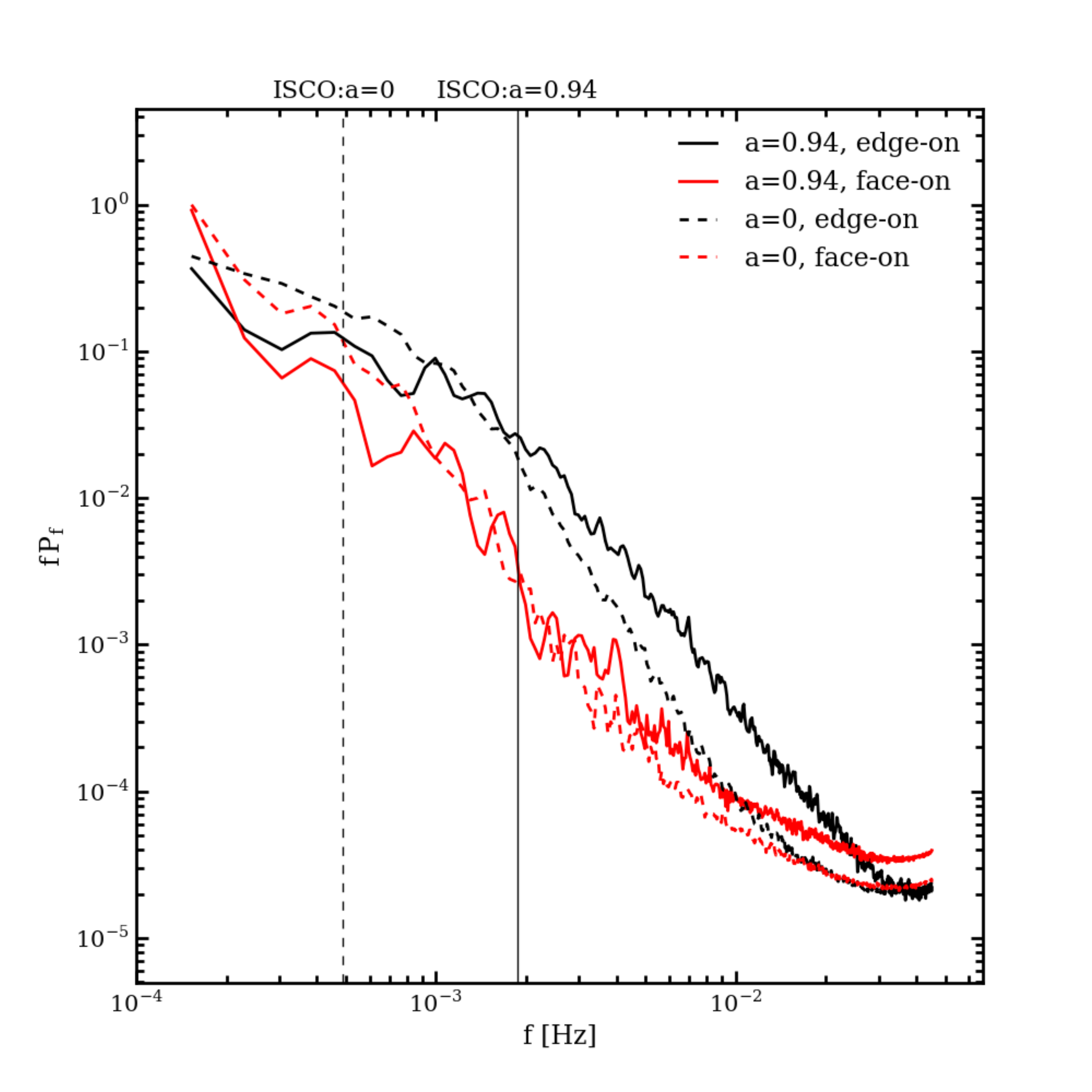}
%\plotone{psd_ftot.pdf}
%\end{center}
\caption[]{PSD of total flux variability for our models.  Solid black,
solid red, broken black, and broken red lines are $a=$0.94 edge-on,
$a=$0.94 face-on, $a=0$ edge-on, and $a=0$ face-on, respectively.
Vertical solid and broken lines indicate ISCO orbital frequency of
$a=$0.94 and $a=0$ black hole, respectively.
The power is normalized by the maximum value of $a=0$ face-on model.}
\label{psd_ftot}
\end{figure}

Figure \ref{psd_ftot} shows the power spectral density (PSD) of the
total flux variability for our four models at $a = 0, a = 0.94$ and
$i = 0^{\circ}, i = 90^{\circ}$.  The vertical lines indicate the ISCO
frequency for $a = 0$ (broken) and $a = 0.94$ (solid).  The total flux
in our models comes from an extended range in radius, so its variability
is not sensitive to disk dynamics at any particular radius.  Overall,
edge-on views have higher power in a wide band around the ISCO frequency
than face-on views (compare black and red lines of the same line type),
while no systematic difference is apparent between different spins with
the same viewing angle (compare the same colors).

All the models exhibit a monotonic decrease in power with frequency.
There is weak evidence for a break close to the ISCO frequency, with
$d\ln P/d\ln f \simeq -2$ below and $d\ln P/d\ln f \gtrsim -3$ above.
Observations \citep{Dexter.et.al.2014} show an $\sim f^{-2}$ spectrum
between $f_{min} \simeq 1.4 {\times} 10^{-3}$Hz and $f_{max} \simeq 3
{\times} 10^{-3}$Hz, with a turnover to a flat spectrum at frequencies
below $\sim 3.5 {\times} 10^{-5}$Hz.  In a future investigation we will
consider the long-timescale behavior and the observed break at $\sim 3.5
{\times} 10^{-5}$Hz by extending the duration of our simulation. The
peak radius of emission in the edge-on views is approximately $5-6M$,
but there is still a large contribution to the total flux from smaller
radii.  The peak emission for face-on views comes from the photon ring
(see the left column in Figure \ref{flux}), but also from a large range
in radius. These highly extended origin of emission for the face-on models
smear out the variability amplitude and lead to an overall lower power
than in edge-on models.  Evidently spatially resolved
structural variability is essential if we want extract more detailed information
about the disk dynamics. However, Figure \ref{psd_ftot} implies the total
flux PSD may allow discrimination between different spin models.

\subsection{Time-Domain Filtering of Image}\label{sec_filtering}

Next, we consider the effects of filtering the mock images in the time
domain.  We first obtain the light curve of each pixel in Figure \ref{flux}
and produce a corresponding PSD following the procedure described in
\S\ref{sec_totalflux}. The PSD is integrated over a frequency band,
and the resulting power is assigned to the same pixel to generate a
time-domain filtered image.  Images reconstructed from observational data will,
of course, have coarser angular resolution. After obtaining the
sets of PSD of the pixels in the ray-traced images, we analyze their
spatial distribution by mapping the power in a band or frequency range
$f_1<f<f_2$; $\mathcal{P}(i,j,f_1,f_2)=\int_{f_1}^{f_2} P(i,j,f)df$
where $i$ and $j$ are the spatial indices of a pixel.

\begin{figure*}
\begin{center}
\hspace*{-10mm}\includegraphics[width=200mm]{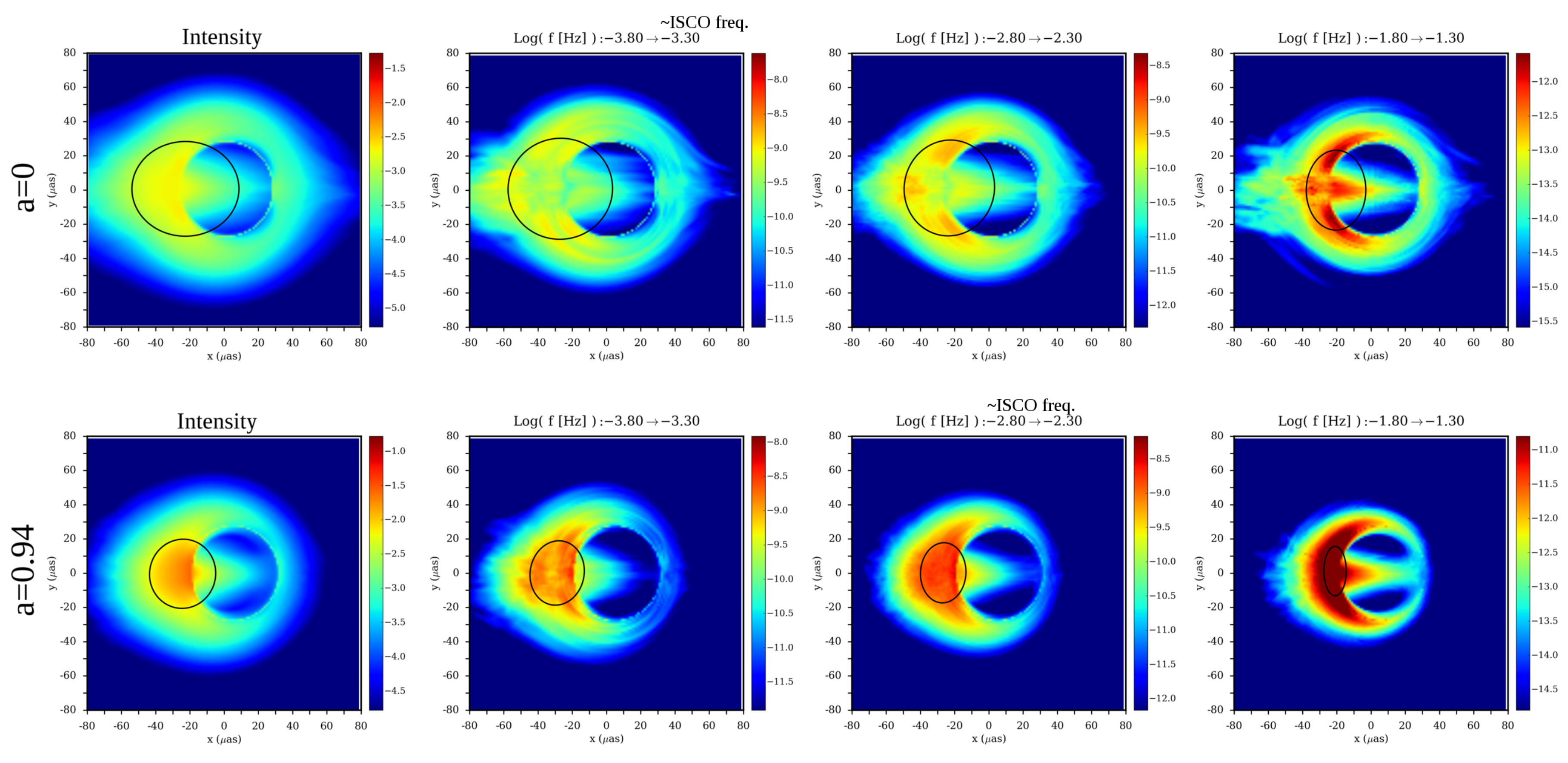}
%\plotone{psp_map_a0_a9_edge.pdf}
\end{center}
\caption[]{Maps of power in time variability for the edge-on view.
The upper and lower rows are for the $a=0$ and $a=0.94$ models,
respectively.  The columns are for different frequency range that the
power is integrated for.  The left most column is for $f=0$, i.e. time
averaged intensity, and from the second left column to the right most
column are, $(f_1,f_2)=(10^{-3.8},10^{-3.3})$, $(10^{-2.8},10^{-2.3})$,
$(10^{-1.8},10^{-1.3})$ in Hz (ISCO orbital frequencies are $10^{-3.3}$Hz
and $10^{-2.7}$Hz for $a=0$ and $a=0.94$, respectively). The color
scale is logarithmic and ranged $\rm{Log}(\rm{AVG}[\mathcal{P}(r<40\mu
as)])\pm 2$ where $r$ is the distance from the centroid.  FOV is
the same as Figure \ref{flux}. The over-plotted ellipses are image
size and orientation found from the covariance matrix described in
\S\ref{sec_filtering}.}
\label{psp_map_edge}
%\end{sidewaysfigure*}
\end{figure*}

\begin{figure*}
\begin{center}
\hspace*{-10mm}\includegraphics[width=200mm]{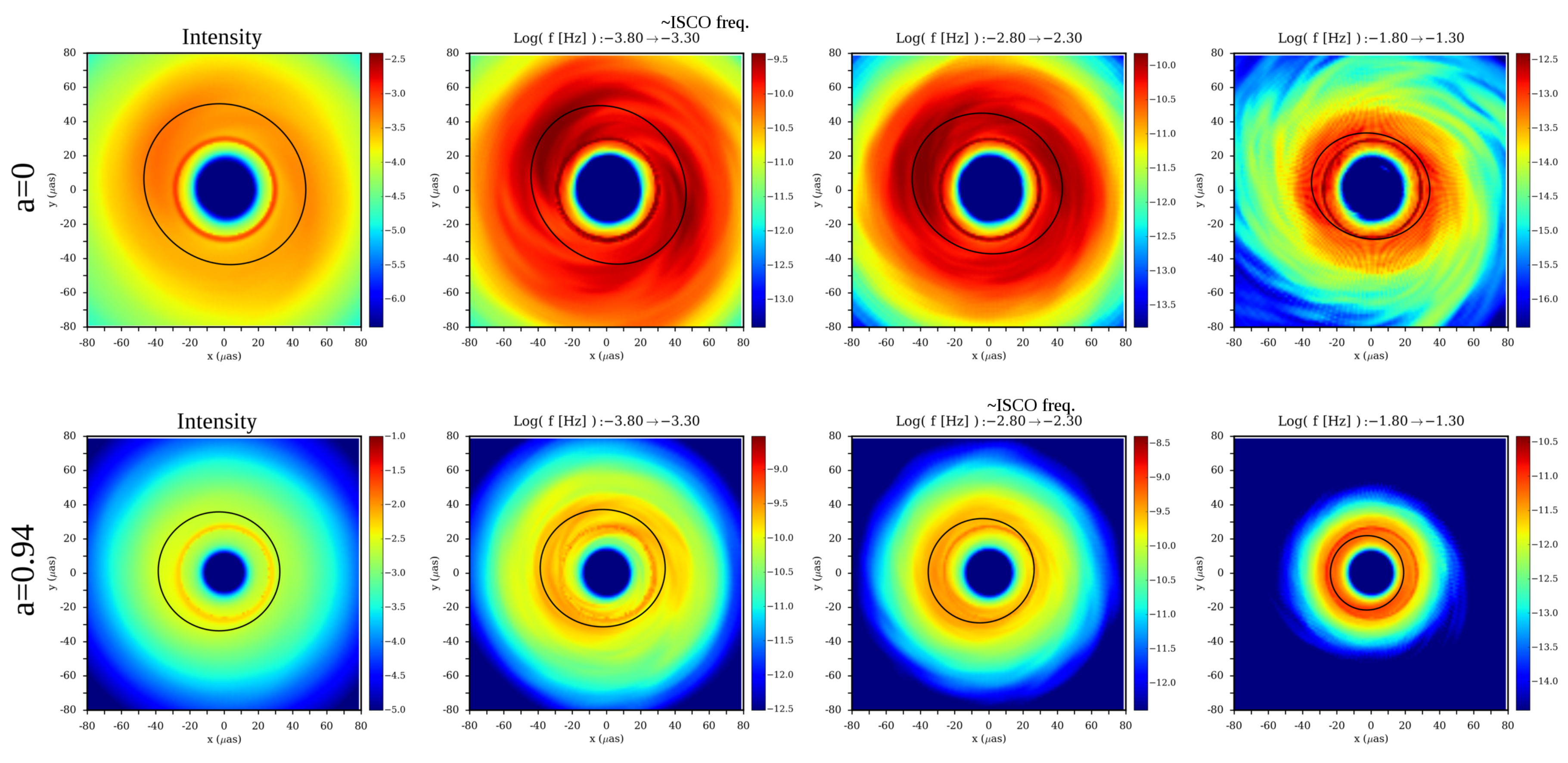}
%\plotone{psp_map_a0_a9_face.pdf}
\end{center}
\caption[]{Same as Figure \ref{psp_map_edge} but for the face-on view.}
\label{psp_map_face}
%\end{sidewaysfigure*}
\end{figure*}

Figures \ref{psp_map_edge} and \ref{psp_map_face} show unnormalized
time-domain filtered images $\mathcal{P}$ for
$(f_1,f_2)=(10^{-3.8},10^{-3.3})$, $(10^{-2.8},10^{-2.3})$,
$(10^{-1.8},10^{-1.3})$ Hz (notice that $f_{\rm{ISCO}}=10^{-2.7}$ and
$10^{-3.3}$ Hz for a=0.94 and 0, respectively) together with the time
averaged intensity map ($f=0$) for the face-on and edge-on views in a
logarithmic color scale. The figures are over-plotted with ellipses to represent the image size
and orientation.  To obtain these ellipses, we first construct
covariance matrix (second moments) of the image pixels and find its
eigenvectors and eigenvalues, $\lambda_1$ and $\lambda_2$.  The
eigenvectors correspond to the principal axes of the images and their
length is the square root of the eigenvalues.  The length of the vectors
is related to the FWHM of the axisymmetric Gaussian model used for the
analysis of VLBI observations by FWHM$/2.3$.  The length of the
principal axes of the ellipses in Figure \ref{psp_map_edge} and
\ref{psp_map_face} is the FWHM, i.e.  $\rm{FWHM}_{1,2} =
2.3\sqrt{\lambda_{1,2}}$.  The eccentricity of the ellipses $\equiv
\sqrt{1-\lambda_2/\lambda_1}$.

%\begin{sidewaysfigure*}

%\begin{sidewaysfigure*}

\begin{figure}[ht!]
%\begin{center}
\hspace*{-0mm}\includegraphics[width=86mm]{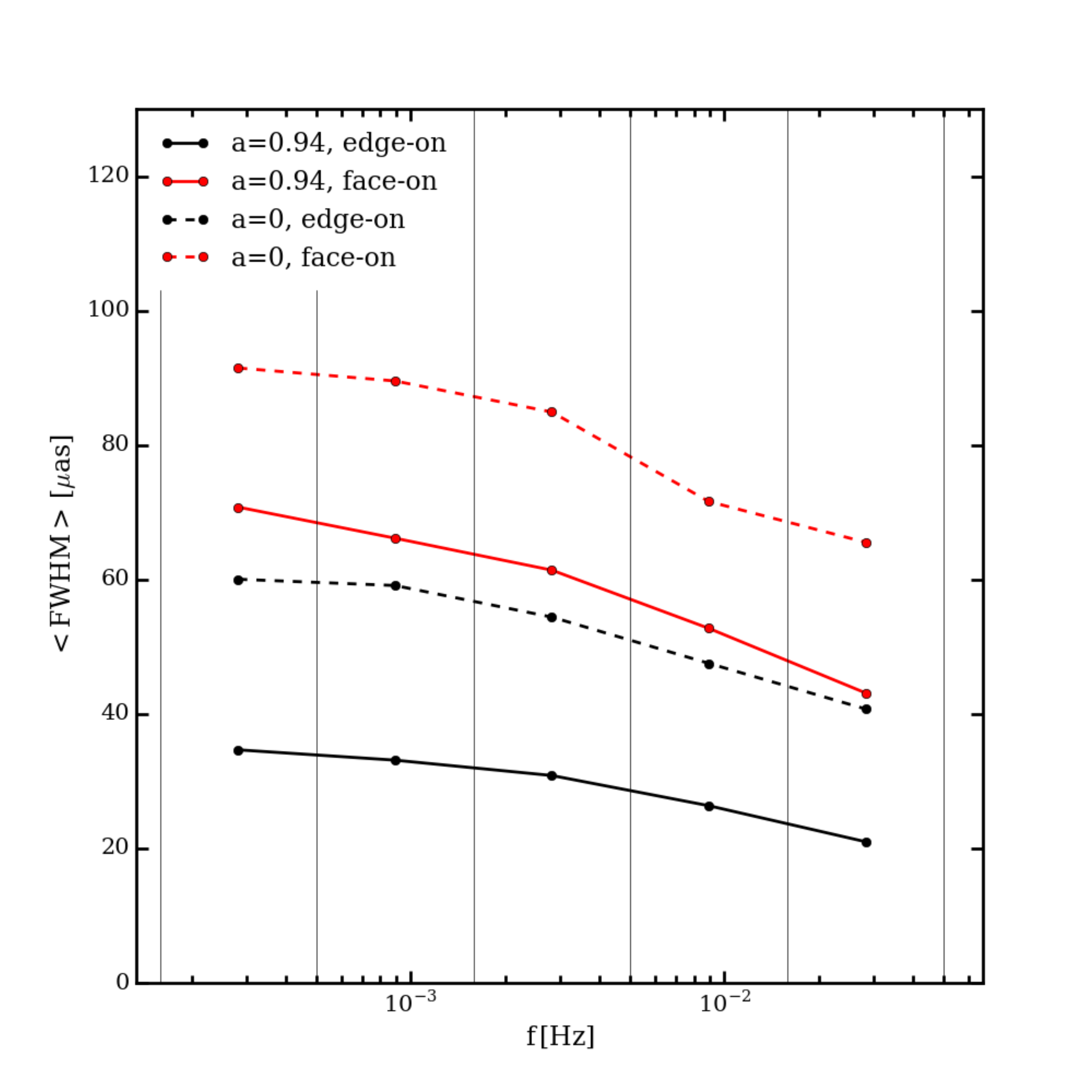}
%\plotone{size.pdf}
%\end{center}
\caption[]{
Filtering frequency versus image size of the time-domain filtered images.
The vertical lines indicate the filtering frequency range for each
plotted points. Solid black, solid red, broken black, and broken red
lines are for $a=0.94$ edge-on, $a=0.94$ face-on, $a=0$ edge-on, and $a=0$
face-on, respectively.}
\label{size}
\end{figure}

The image size decreases as the filtering frequency increases for all
the models in Figures \ref{psp_map_edge} and \ref{psp_map_face}.  In
general, the smaller $r_{emiss}$, the higher
the power at high filtering frequencies and the lower the power at
low filtering frequencies, if the intensity is fixed.  This naturally
arises because the high orbital frequency at the inner radii of the disk
produces rapid variability.  Emission from large radii has relatively
higher power at low filtering frequencies, and hence the image size
is larger due to the extended area of emission from outer radii
(see Figure \ref{emiss_r}).  On the other hand, emission from small
radii occupies a smaller region in the images and the image size gets
smaller for high filtering frequencies.  This result confirms that
our filtering technique is able to visualize a spatial distribution of
the local characteristic timescale at $r_{emiss}$.
Figure \ref{size} shows the relation
between the image size and the filtering frequency range for all
the cases where the ``size'' is the average of FWHM$_1$ and FWHM$_2$.
The size decreases with increased filter frequency, and ranges from 30\%
to 50\% depending on the model considered.

\begin{figure}[ht!]
%\begin{center}
%\plotone[width=160mm]{ecc.pdf}
\hspace*{-0mm}\includegraphics[width=86mm]{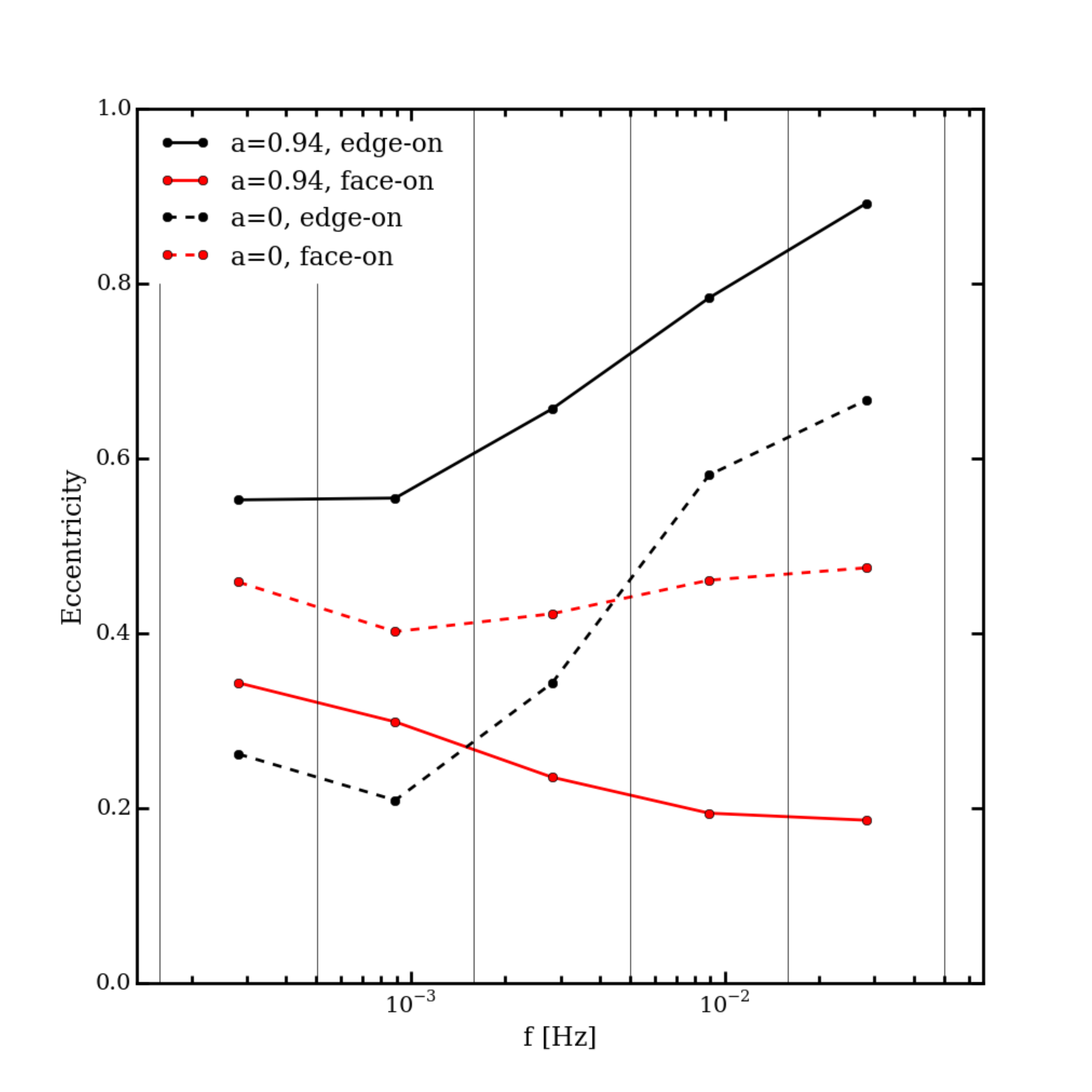}
%\plotone{ecc.pdf}
%\end{center}
\caption[]{Same as Figure \ref{size} but for the eccentricity of the filtered images.}
\label{ecc}
\end{figure}

The high frequency filtered edge-on images (rightmost column in Figure
\ref{psp_map_edge}) nicely traces the distribution of small emission
radii (see Figure \ref{emiss_r}) for both of the high spin models.
This suggests that image eccentricity has a dependence on black
hole spin, especially at high filtering frequencies.  The eccentricity
of the edge-on filtered images in general increases with the filtering
frequency, and the high-spin model has higher eccentricity at all the
filtering frequencies than the low-spin model; see Figure \ref{ecc} for
the dependence of eccentricity on the filtering frequency.  Unlike the
edge-on views, eccentricity in the face-on images does not depend much
on the filtering frequency. In principle, the degeneracy of spin and
viewing angle can be avoided by the unique dependence of eccentricity on
the filtering frequency. Surprisingly, face-on images are not always
very ``round", as we see the face-on $a=0$ model in Figure \ref{ecc}
has eccentricity $>0.4$ due to the presence of slowly evolving
non-axisymmetric structures.

The time-domain filtering technique highlights the direction of the disk's
orbital axis, as the semi-major axis of the ellipses are aligned with it.
Even if the black hole shadow is not well resolved or apparent in time-averaged
data, filtering at high frequency would show anti-symmetric properties
in the variable structure as long as the disk is not viewed face-on.
We note that our technique might be used to test if an
observed time-averaged feature is a true black hole shadow: if it is,
then a similar feature should grow more prominent as the filtering 
frequency increases.
%HOTAKA: please check that this is correct.

\subsection{Application to EHT Data}\label{sec_filtering_real}

Real EHT data is sparsely sampled in the spatial Fourier domain (U-V plane)
and has superposed atmospheric and thermal
noise as well as interstellar scattering.  Thus reconstruction of even static 
images require sophisticated approaches \citep[e.g.][]{Bouman.et.al.2016,Chael.et.al.2016}.
Nevertheless, reconstruction of high time resolution
images from incomplete UV-coverage is possible using several newly
developed techniques (Johnson et al. 2017, submitted).

We performed a preliminary test to apply the time-domain filtering technique to time
variable images produced by a synthetic observation of our high spin
edge-on view model, using the array configuration for the EHT 2017 
campaign. However, the effects of interstellar scattering
and realistic, irregular time sampling (i.e. due to telescopes off-source f
or phase calibration) were not included.  We found that the technique could 
reproduce the trend the image size increases and eccentricity decreases as the filtering
frequency increases.
%To sum up, although this paper uses real-space
%images for demonstrating the time-domain filtering technique, 
%the same technique could be employed to data in the UV plane.}

\subsection{Slow Light and Fast Light}

Here, we investigate the effect of using a ``fast light"
approximation by comparing fast light and slow light time-domain
filtered images. Implementation of ``slow light" in General Relativistic
Ray-tracing code has been done by several researchers \citep[e.g.][Jason
Dexter, personal communication]{Dolence.et.al.2012} and it is known to
have a significant effects on high frequency variability (Chi-kwan Chan,
Thomas Bronzwaer, personal communication).

\begin{figure}[ht!]
%\begin{center}
\hspace*{-5mm}\includegraphics[width=95mm]{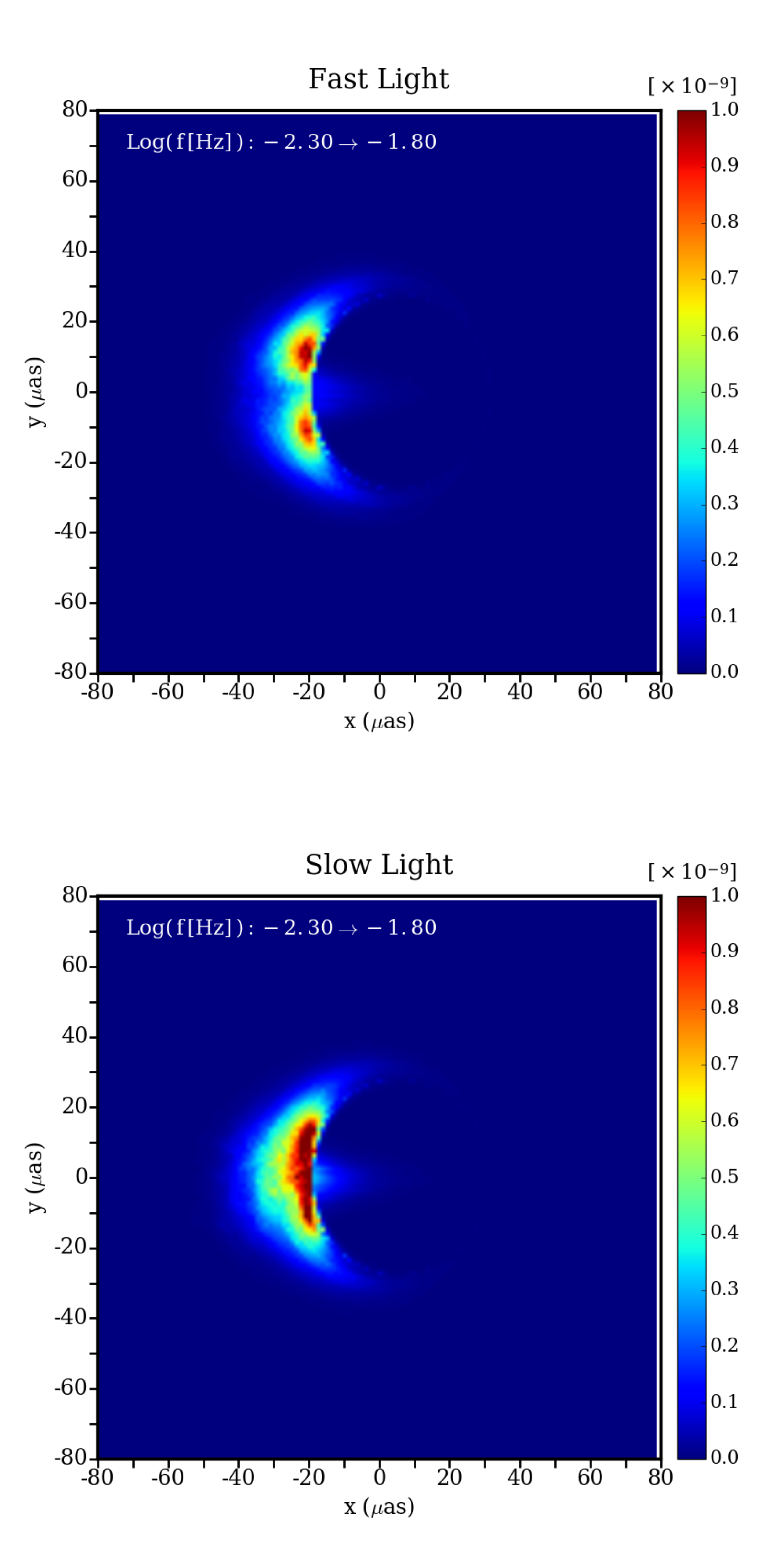}
%\plotone{slowfast.pdf}
%\end{center}
\caption[]{Comparison of the time-domain filtered images calculated using the fast light approximation
(upper) and without the approximation (lower). The selected images are for the $a=0.94$
edge-on view model with the filtering frequency range $10^{-2.3}-10^{-1.8}$Hz (ISCO orbital
frequency is $\sim 10^{-2.7}$Hz). The color is linearly scaled.}
\label{slowfast}
\end{figure}

We found the approximation introduces non-negligible effects on filtered
image morphology for the edge-on view, at the high filter frequencies.
As an example, Figure \ref{slowfast} shows a comparison of the time
 filtered images with and without the fast light approximation where
 the properties of the images are the $a=0.94$ model, edge-on view,
and filter frequency range of $10^{-2.3}-10^{-1.8}$Hz (same as the
lower row, third column panel in Figure \ref{psp_map_edge} but with a
 linear color scale). The power around the equatorial plane in the
fast light image is approximately 50\% less than in the corresponding
 slow light image. Evidently, the use of slow light is essential for
our time-domain filtering technique.

For the same model, the effect of the approximation on the time-domain
filtered image is much less apparent at the low filter frequencies,
approximately below ISCO orbital frequency $\sim 1.9 \times 10^{-3} =
10^{-2.7}$Hz. The total flux light curve of the slow light calculation
is slightly smoother than that of the fast light but no qualitative
difference is apparent.

%The fast light filtered images consistently has higher image size and eccentricity than the slow
%light images for our selected filter frequency ranges between $10^{-3.8}$ and $10^{-1.3}$ but only by few-10\%.

%%%%%%%%%%%%%%%%%%%%%%%%%%%%%%%%%%%%%%%%%%%%%%%%%%%%%%%%%%%%%%%%%%%%%%%%%%%%

\section{Turbulence and Observed Variability}\label{sec_turbulence}

%Now we turn to study the relation between the observed light curve for a
%single pixel and the dynamics of its emission point, $\vec{x}_{emiss}$.
%We measure time variability of the emissivity at an emission point in the
%disk's equatorial plane, and then compare this to the variability of the flux
%measured in the corresponding pixel.  The PSD of the emissivity is obtained
%using the same method as for the PSD of the flux.

%[HS: could you add a reference to Ryan et al. 2017, our new
%paper on shearing box convergence?]
While numerous simulations of magnetized turbulent
accretion disks have been conducted, the resultant turbulence
structure depends on the numerical resolution, the initial
condition/setting of the simulation, and the physics included in the model
\citep[e.g.][]{Hawley.et.al.2011,Shiokawa.et.al.2012,Kunz.et.al.2016,Shi.et.al.2016,Ryan.et.al.2017}.
In the absence of clear theoretical guidance, then, it is useful to understand
what information time-variable imaging might contain about the structure
of turbulence in Sgr A*'s accretion flow.   Spatially resolved
time-domain data retains information about location specific disk dynamics,
which is inaccessible in total flux light curves
of Sgr A* that average emission from a broad range of radii in the disk.

Here, we study the relation between the observed light curve for a
single pixel and the dynamics of its emission point, $\vec{x}_{emiss}$,
to explore how the disk structure at the emission point is translated
into the observed variability. We measure time variability of the
emissivity at an emission point in the disk's equatorial plane, obtain
its PSD, and then compare this to the PSD obtained from variability of
the flux measured in the corresponding pixel.  The PSD of the emissivity
is obtained using the same method as for the PSD of the flux.

\begin{figure*}[ht!]
%\begin{center}
\hspace*{-4mm}\includegraphics[width=185mm]{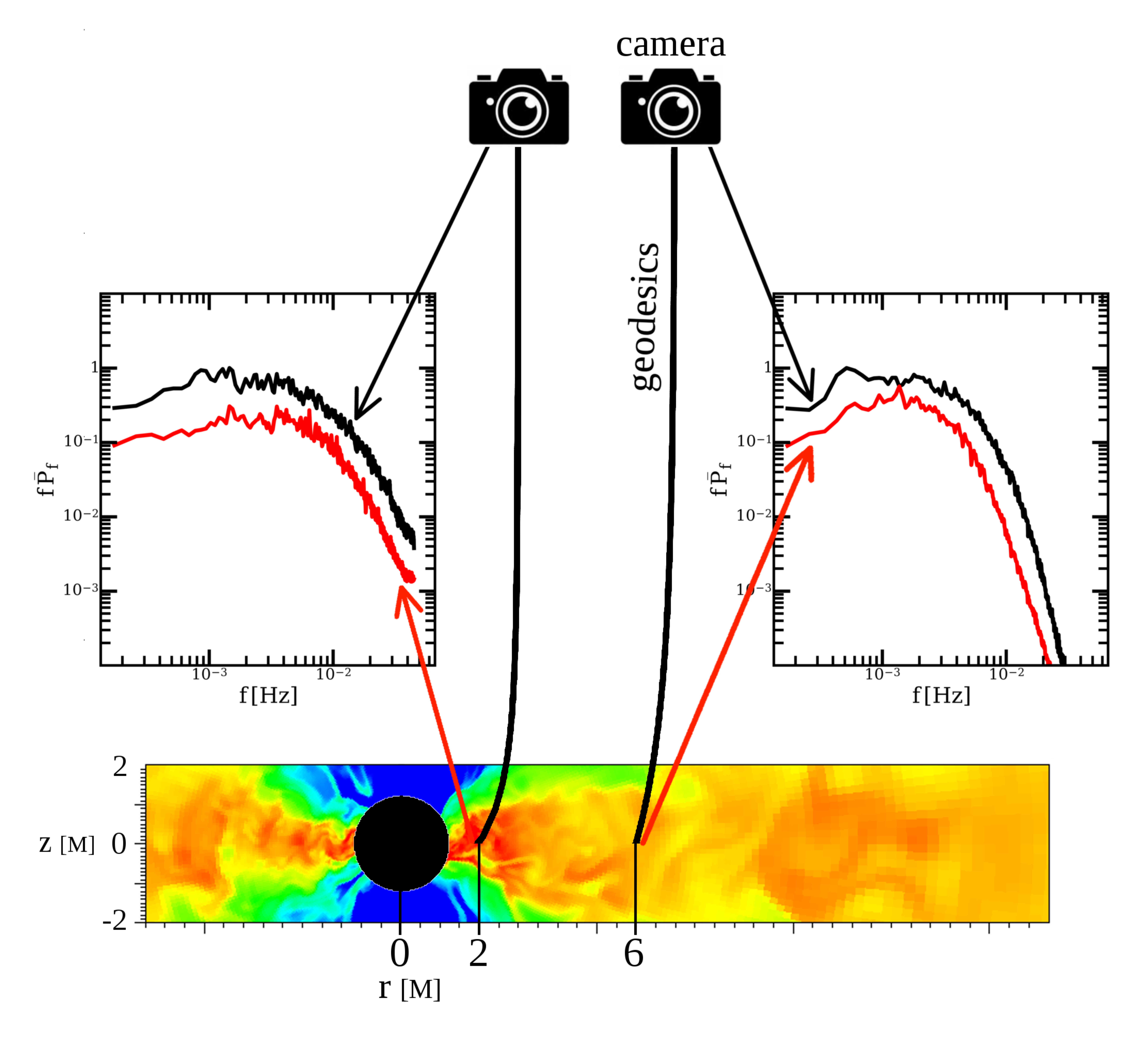}
%\plotone{indisk.pdf}
%\end{center}
\caption[]{Comparison of observed PSDs in pixels and PSDs of temporal emissivity
fluctuation in time measured at the pixels' emission points
$\vec{x}_{emiss}$ in the disk. The selected examples are for pixels
that have emission radius $r_{emiss}=2M$ (the left panel) and $6M$
(the right panel) in the equatorial plane for the face-on, $a=0.94$
model. The black PSD curves pointed by the black arrows from the cameras
(represent the pixels) are the observed PSDs in the selected pixels, and
the red PSD curves pointed by the red arrows from the emission points
are the PSDs of emissivity fluctuation measured at those points.
The normalizations offset to make both curves visible.
Emissivity is average value of 9 grids (3 in radial and 3 in poloidal
direction) adjacent to a grid at the emission points. The pseudo-color
plot shows density distribution in an example azimuthal and time slice of
the accretion disk. The color is logarithmically scaled.
%Note the PSDs
%are the normalized PSDs ($f\bar{P}(f)$, see equation \ref{psd_norm}),
%and further normalized by the maximum value of the observed PSDs in
%each panel.
}
\label{indisk}
\end{figure*}

%[HS: this is a cool figure.  Why are the PSDs different in amplitude?
%how do you normalize the PSD?  I would think there would be comparatively
%less fluctuation in the flux than in the emissivity.]
Figure \ref{indisk} presents PSDs of observed light curves in selected
pixels for the face-on $a=$0.94 model, and PSDs of emissivity fluctuations
measured at the pixels' emission points. The selected pixels
have their emission radii $r_{emiss}=2M$ and $6M$ in the equatorial
plane. Although there are differences in detail between the emissivity
and flux PSDs, it is evident that they have remarkably similar shapes, consistent
with our suggestion that variability in an image pixel and dynamics
at the emission point associated with that region of the image are
closely related. This correlation is retained even for pixels
that have an extended emission region in the disk such as the brightest
region in the edge-on images.

%What does the observed PSD tell us about the disk dynamics?  The
%temporal PSD of emissivity at a fixed location in the disk is directly
%connected to the PSD of spatial fluctuation of the emissivity in azimuthal
%direction, since much of the variability is generated by slowly evolving
%structures that are dragged with the plasma as it orbits the hole.
%Therefore, the characteristic temporal frequencies of the variability
%are directly related to the orbital frequency of the disk:
%\begin{equation}\label{EQ_mode}
%f_c(r)=m\Omega_p(r)\sim 2\pi m f_K(r)\text{,}
%\end{equation}
%where $\Omega_p$ is the ``pattern speed'' of the emissivity which
%(consistent with our claim that these are generated by slowly evolving
%structures) we approximate as the Keplerian orbital velocity
%$f_K(r)=[2\pi(r^{3/2}+a)]^{-1}$.  Here $m$ is the mode number, i.e. the
%emissivity has a characteristic angular structure $e^{i m\phi}$.

What does the observed PSD tell us about the disk dynamics? 
The turbulent velocities in the disk are small compared to the orbital
speed.  This suggests a picture in which the disk fluctuations are
regarded as frozen in, and dragged across the line of sight by
orbital motion.   This implies that the temporal PSD of
emissivity at the location is directly connected to the PSD of spatial
fluctuation of the emissivity in azimuthal direction.  The dominant
variability frequency should correspond to the time required for a
characteristic turbulent structure to orbit across the line of sight.
The characteristic azimuthal size of turbulent structures
in the disk is the azimuthal correlation angle $\lambda_{\phi} \ll 2\pi$
\citep{Shiokawa.et.al.2012}, which corresponds to an azimuthal model
number $m = 2\pi/\lambda$.  Therefore 
\begin{equation}\label{EQ_mode}
f_c(r) \simeq m \frac{\Omega_p(r)}{2\pi} \simeq m f_K(r)\text{,}
\end{equation}
where $\Omega_p$ is the pattern speed of the emissivity, which
we assume is approximately the Keplerian orbital velocity 
$f_K(r)=[2\pi(r^{3/2}+a)]^{-1}$.

The angular correlation length $\lambda_{\phi}$ of the emissivity in
a turbulent disk could be estimated based on the preceding argument.
The characteristic mode frequency for the highest mode number
$m=2\pi/\lambda_{\phi}$ appears as a break in the PSD; higher
frequency fluctuations are weaker because there is comparatively
little corresponding small-scale structure in the disk.  It turns out
that there is a mild break in most of the observed pixel PSDs, 
including the PSDs in Figure \ref{indisk}, and these breaks correspond
to a mode number $m=8-10$, consistent with \citet{Shiokawa.et.al.2012}.

%[HS: we should discuss this - I'm not sure what you're trying to say.]
%Application of our analysis to a time-filtered image obtained in a real EHT observation
%is either finding emission radius distribution in the image from a
%plausible turbulent disk model or finding the turbulence correlation length
%from foreknown emission radius distribution that would be
%estimated from further parameter surveys of Sgr A* \citep[former works
%can be found in e.g.][]{Moscibrodzka.et.al.2014,Chan.et.al.2015b}.
%The obtained correlation length could be even shorter than the highest
%spatial resolution available for the current EHT.

Note that the azimuthal structure of the emissivity strongly on
wavelength, as does the synchrotron emissivity \citep{Leung.et.al.2011}.
At shorter wavelength, e.g. in the near infrared, the
azimuthal profile is dominated by a few emission spikes rather than by
the more smoothly distributed turbulent fluctuations that dominate in
the submillimeter. Such spikes could cause Quasi Periodic Oscillations 
(QPOs) at close to the orbital frequency \citep{Dolence.et.al.2012}.

%%%%%%%%%%%%%%%%%%%%%%%%%%%%%%%%%%%%%%%%%%%%%%%%%%%%%%%%%%%%%%%%%%%%%%%%%%%%

\section{Summary}

The Event Horizon Telescope project is aimed at spatially resolving
 the mm/submm emission from Sgr A* on scales of a few Schwarzschild
radii. Turbulence in the accretion flow surrounding Sgr A* is expected
to cause time variability in the sky brightness that can be measured
and potentially resolved by EHT.
In this paper, we have introduced  a time-domain filtering technique
that filters time variable images into arbitrary temporal
frequency ranges. We have demonstrated the technique's potential by
exploring the connection between the accretion flow and image variability
through the use of mock, time-dependent, 1.3mm images based on GRMHD
simulations.
%In this paper, we have explored this connection through
%the use of mock, time-dependent, 1.3mm images based on GRMHD simulations.

%While full temporal power spectra of individual image pixels may not be measurable
%by EHT, it may be possible to extract reasonable signal-to-noise images
%of power in broad frequency bands.   We have therefore explored time-domain
%filtering of our time-dependent mock images.
By quantifying the image properties using image moments, we find that
the filtered images depend strongly on the filter frequency, and make
the following predictions:

\begin{itemize}
\item[$\bullet$]   Emission from small radii in the disk has rapid
variability because it originates from the fastest moving turbulent
structures.  It appears as relatively concentrated structure in the
images and hence the image size at the high filter frequencies are
smaller than those low frequencies.  The image size decreases with 
increasing filter frequency for all the
parameters we explore ($a=0$ and 0.94, face-on and edge-on viewing
angles) by 30-40\% for the possible temporal frequency range in the real
VLBI observations.

\item[$\bullet$] High frequency filters pick out the photon
ring, which is built up from emission close to the
photon orbit.  Although the size of the photon ring is insensitive to spin,
its characteristic variability frequencies are spin dependent because
the emission radius (photon orbit radius) is also spin-dependent.  

\item[$\bullet$] The filtered image's eccentricity (a measure of the ellipticity
of the image) is spin-dependent in disks seen edge-on.   Our models 
show the edge-on $a=$0.94 model images are more eccentric than 
the edge-on $a=$0 model at all filter frequencies.

\item[$\bullet$] The filtered image's major axis is aligned with the
spin axis of the black hole and disk\footnote{In the models considered
here, the disk angular momentum is aligned with the black hole spin}.   
The spin axis can therefore be accessed using our filtering technique.

\item[$\bullet$] The filtered image's eccentricity is less affected by filter frequency
in face-on views. This can be used to differentiate the edge-on and
face-on views.
\end{itemize}

The dependence of the image morphology on filter frequency is a direct
consequence of the coupling of disk dynamics to variability.  By
comparing PSDs, we showed that the flux variability is tightly
correlated with the emissivity fluctuations at the point of peak
emission along the line of sight.  The
predominant source of the variability is slowly evolving turbulence
structure passing through the emission point at the orbital speed. Therefore,
the observed temporal PSD in a pixel is reflection of the
spatial PSD of the emissivity turbulent structure at the emission
point. This indicates that spatial structure of disk turbulence,
and moreover the emission radius distribution in the observational
images, are accessible through image variability.

%Our time filtering technique demonstrates that it is possible in
%principle to distinguish spin, viewing angle, and direction of spin axis
%of different models using only movies of the lensed emission.  It is as
%yet unclear how well this can be done using real EHT data; we plan to
%explore this in a future publication, that will also improve on the
%current effort by using more realistic simulations that permit 
%$T_p/T_e$ to vary in time and space.

The time filtering approach described here shows that it is possible
to directly study black hole spin, viewing angle and spin axis of the
Sgr A* system using movies of lensed emission produced by GRMHD simulations.
Next steps in this area will include development of algorithms tailored
to EHT data sets that aim to estimate these fundamental parameters.
Parallel extension of GRMHD simulations to include other effects such
as variations in electron/proton temperature ratio in the accretion flow
will be useful in application of these techniques to other EHT targets
such as M87.

This work was supported by a grant from the National Science Foundation
(NSF; AST-1440254) and through an award from the Gordon and Betty Moore
Foundation (GBMF-3561). C.F.G. was supported by NSF grant AST-1333612,
a Simons Fellowship, and a visiting fellowship at All Souls College,
Oxford.  C.F.G is also grateful to Oxford Astrophysics for their
hospitality. The authors would like to acknowledge Scott Noble for
providing GRMHD code HARM3D. We thank Avi Loeb, Pierre Christian,
and Michael Johnson for fruitful discussions, and Andrew Chael for
assistance in editing the paper.

\bibliography{ms}
\end{document}